\documentclass[aps,twocolumn]{revtex4-1}
\usepackage{amsmath,amssymb}
\usepackage{slashed}
\usepackage{graphicx}
\usepackage[utf8]{inputenc}
\usepackage{tikz}
\usetikzlibrary{shapes.geometric, decorations.pathmorphing, decorations.markings, arrows.meta}
\usepackage{xcolor}
\usepackage{bm}
\usepackage[T1]{fontenc}
\usepackage{gauss} 
\usepackage[top=1.0in,bottom=1.0in,left=1.0in,right=1.0in]{geometry}
\usepackage[colorlinks,linkcolor=blue,citecolor=blue]{hyperref}
\usepackage{subfig}
\newcommand{\be}{\begin{equation}}
\newcommand{\ee}{\end{equation}}
\newcommand{\bea}{\begin{eqnarray}}
\newcommand{\eea}{\end{eqnarray}}
\newcommand{\ba}{\begin{eqnarray}}
\newcommand{\ea}{\end{eqnarray}}

\def\be{\begin{eqnarray}}
\def\ee{\end{eqnarray}}
\def\bea{\be}
\def\eea{\ee}

\def\roughly#1{\mathrel{\raise.3ex\hbox{$#1$\kern-.75em%
\lower1ex\hbox{$\sim$}}}}

\begin{document}

%\title{Deeply virtual Compton scattering  in holographic QCD: \\
%Quark and Gluon GPDs at non-zero skewness}

%\title{GPDs from holographic moment parametrization at finite skewness: \\a comparison to lattice QCD}

\title{Parametrization of GPDs from t-channel string exchange in AdS spaces}

\author{Kiminad A. Mamo}
\email{kamamo@wm.edu}
\affiliation{
Physics Department, William \& Mary, Williamsburg, VA 23187, USA}

\author{Ismail Zahed}
\email{ismail.zahed@stonybrook.edu}
\affiliation{Center for Nuclear Theory, Department of Physics and Astronomy, Stony Brook University, Stony Brook, New York 11794-3800, USA}

\date{\today}

\begin{abstract}
We introduce a string-based parametrization for nucleon quark and gluon generalized parton distributions (GPDs) that is valid for all skewness. Our approach leverages conformal moments, representing them as the sum of spin-j nucleon A-form factor and skewness-dependent spin-j nucleon D-form factor, derived from t-channel string exchange in AdS spaces consistent with Lorentz invariance and unitarity. This model-independent framework, satisfying the polynomiality condition due to Lorentz invariance, uses Mellin moments from empirical data to estimate these form factors. With just five Regge slope parameters, our method accurately produces various nucleon quark GPD types and symmetric nucleon gluon GPDs through pertinent Mellin-Barnes integrals. Our isovector nucleon quark GPD is in agreement with existing lattice data, promising to improve the empirical extraction and global analysis of nucleon GPDs in exclusive processes, by avoiding the deconvolution problem at any skewness, for the first time.
\end{abstract}

\maketitle

\textit{\bf Introduction}.—The structure of the nucleon has been under scrutinity for over half a century, and for a good reason. Nucleons are the key building blocks in the composition of most visible matter in our universe. Many dedicated experiments have now established that the sub-constituents of the nucleon are quarks and gluons, as described by quantum chromodynamics (QCD), the theory of strong interactions in the standard model. 

The  massless gluons and nearly massless quarks, are at the origin of most visible mass, a remarkable feat of dimensional transmutation in QCD. This
mass from no mass, has been checked numerically by dedicated QCD
lattice simulations. Yet, quarks and gluons or their wee partonic manifestation on the light front,  cannot be singled out for dedicated scrutinity, due to confinement in QCD.

The partonic structure of most hadrons is probed via hard inclusive and semi-inclusive processes.  Due to relativistic time dilation and asymptotic freedom, hard processes can be separated into perturbative and non-perturbative elements like parton distribution functions (PDFs) and fragmentation functions (FFs)~\cite{Callan:1969uq,Gross:1973id,Politzer:1973fx,Collins:1989gx}.

Central to this study are Generalized Parton Distributions (GPDs), which are extended PDFs represented as off-forward matrix elements of leading twist QCD operators in boosted hadron states~\cite{Muller:1994ses,Radyushkin:1996nd,Radyushkin:1996ru,Ji:1996nm,Radyushkin:1997ki,Ji:1998pc,Goeke:2001tz}, see~\cite{Diehl:2003ny,Belitsky:2005qn,Mezrag:2022pqk} for review. These multidimensional distributions provide a spatial imaging of partons in the nucleon, with insights into the nucleon charge, spin, mass, and mechanical structure~\cite{Diehl:2003ny,Belitsky:2005qn,Mezrag:2022pqk}.
Advances in lattice QCD are starting to address  some of these issues~\cite{Lin:2020rxa,Lin:2021brq,Alexandrou:2020zbe,Alexandrou:2021bbo,Bhattacharya:2022aob}, using the theoretical approaches proposed in~\cite{Ji:2013dva,Ji:2014gla,Ji:2020ect} and~\cite{Radyushkin:2017cyf,Orginos:2017kos}. 
For completeness, we note that a number of phenomenological models have addressed the quark GPDs by directly parametrizing them in $x$ (momentum fraction) space \cite{Ji:1997gm,Anikin:2001zv,Scopetta:2002xq,Boffi:2002yy,Scopetta:2003et,Ji:1997gm,Anikin:2001zv,Scopetta:2002xq,Boffi:2002yy,Scopetta:2003et,Petrov:1998kf,Penttinen:1999th,Goeke:2001tz,Vega:2010ns,Rinaldi:2017roc,Traini:2016jko,deTeramond:2018ecg,Mondal:2015uha,Chakrabarti:2015ama,Mondal:2017wbf,Chakrabarti:2013gra,Gurjar:2022jkx,Shuryak:2023siq,Liu:2024umn,Radyushkin:1998es,Radyushkin:1998bz,Polyakov:1999gs,Musatov:1999xp,Goloskokov:2009ia,Goloskokov:2005sd,Goloskokov:2007nt,Goeke:2001tz,Vanderhaeghen:1998uc,Vanderhaeghen:1999xj,Guidal:2004nd}, by parametrizing their conformal moments at small and zero skewness \cite{Polyakov:1998ze,Polyakov:2002wz,Mueller:2005ed,Kumericki:2007sa,Kumericki:2009uq,Polyakov:2009xir,Muller:2014wxa,Cuic:2024fum,Guo:2022upw,Guo:2023ahv,Cuic:2023mki}, and more recently using lattice data for Compton form factors~\cite{Hannaford-Gunn:2024aix}.

The GPDs are accessed experimentally through exclusive processes. Particular  efforts are currently underway to further elucidate GPD phenomenology~\cite{dHose:2016mda,Guidal:2013rya,Kumericki:2016ehc}. A number of dedicated experiments are being designed at major facilities to probe them, such as COMPASS at CERN, STAR at RHIC and in various halls at JLab, with  major contributions slated from the upcoming high luminosity electron ion collider (EIC) facility~\cite{AbdulKhalek:2021gbh,Anderle:2021wcy}.

GPDs can be extracted from deep virtual Compton scattering (DVCS), and deep virtual meson production (DVMP) processes. However, the non-uniqueness in the deconvolution of GPDs from the Compton Form Factors of these processes, especially when the GPDs are parametrized directly in the momentum fraction \( x \) space using double distributions \cite{Radyushkin:1998es,Radyushkin:1998bz}, presents a significant theoretical challenge for their model-independent extraction from data~\cite{Guidal_2013,Kumeri_ki_2016,Bertone:2021yyz,Moffat:2023svr}. The way to bypass the deconvolution problem is by parametrizing the conformal moments of GPDs \cite{Guo:2022upw,Guo:2023ahv}, which are highly constrained by Lorentz invariance (polynomiality condition), and have a dual interpretation in terms of t-channel spin-j resonances \cite{Polyakov:1998ze,Polyakov:2002wz}.

%Earlier attempts can be found in~\cite{Ademollo:1969wd,Gatto:1972gpd,Schierholz:1973nw,Schierholz:1974pz}.

The holographic approach to QCD in the double limit of a large number of colors and strong 't Hooft coupling, has emerged as a powerful method for addressing 
non-perturbative phenomena in QCD, using dual gravity~\cite{Nastase:2015wjb} (and references therein). When addressing QCD scattering processes in AdS spaces, holography embodies the old dual string 
model  with all its phenomenological successes~\cite{Veneziano:1974dr,Frampton:1986wv}. It is the most economical way of enforcing QCD symmetries, duality and crossing symmetries,  within a well defined and minimal organisational principle using field theory consistent with Lorentz invariance and unitarity. 

In this letter, we propose a holographic string-based approach for analyzing GPDs,
thereby bypassing the deconvolution problem~\cite{Guidal_2013,Kumeri_ki_2016,Bertone:2021yyz,Moffat:2023svr}.
%in the ERBL and DGLAP regime, employing Gegenbauer moments and the Sommerfeld-Watson transform to maintain analyticity across all skewness regions. 
Our approach streamlines the QCD analysis by using fewer parameters, and circumvents the deconvolution problem through universal, model-independent and unique string-based parametrizations of the conformal moments of GPDs consistent with the polynomiality condition at any skewness $\eta$, for the very first time.
%%%%

\textit{\bf Quark and gluon GPDs}.—GPDs are bilocal quark and gluon correlators 
on the light front~\cite{Belitsky:2005qn} (and references therein). Their physical content can be split into the large-x region $\eta\leq x\leq 1$ (DGLAP), and the low-x region $0\leq x\leq \eta$ (ERBL).
Here  $x$ is the parton fraction of the struck quark, and $\eta$ its longitudinal skewness, i.e. the fraction of light cone momentum transfer to the nucleon, for fixed  4-momentum transfer $\sqrt{-t}$.

The GPDs in the ERBL regime  are  organized in a series expansion in conformal partial waves (PWs)  represented by Gegenbauer polynomials in parton-x~\cite{Belitsky:2005qn}. These PWs encode the dynamics of quarks and gluons, bridging parton distribution functions and form factors. The series is analytically continued into the DGLAP  using the Mellin-Barnes integral, which results in GPDs expressed as integral transforms of spin-j conformal moments $\mathbb{F}_q(n, \eta, t; \mu)$ at fixed resolution $\mu$. 

Specifically, the singlet (+) and non-singlet ($-$) combinations of quark GPDs, 
are given in terms of the Mellin-Barnes integrals of the conformal moments~\cite{Mueller:2005ed}
%denoted as $H^{(\pm)}_q(x, \eta, t; \mu)$, are given through the anti-symmetric and symmetric Sommerfeld-Watson transformations, or equivalently
% Mellin-Barnes integrals
\bea
\label{F(x,eta,t)-MBValenceVector}
&&H^{(\pm)}_q(x, \eta, t; \mu) = \nonumber\\
&&\frac{1}{2i}\int_{\mathbb{C}} dj\, \frac{1}{\sin(\pi j)} (p_j(x,\eta) \mp p_j(-x,\eta)) \mathbb{F}_q^{(\pm)}(j, \eta, t; \mu),\nonumber\\
\eea
also known as the Sommerfeld-Watson transformations, with the quark PWs~\cite{Mueller:2005ed}
\bea
\label{Def-p-allhere}
&&p_j(x,\eta) = \nonumber\\
&&\theta(\eta - |x|) \frac{1}{\eta^j} \mathcal{P}_j\left(\frac{x}{\eta}\right) + \theta(x - \eta) \frac{1}{x^j} \mathcal{Q}_j\left(\frac{x}{\eta}\right)
%\,,\nonumber\\
\label{Def-p-P}
\eea
with
\bea
&&\mathcal{P}_j\left(\frac{x}{\eta}\right) = \frac{2^{j}\Gamma(3/2+j)}{\Gamma(1/2)\Gamma(j)}\nonumber\\
&&\times\left(1+\frac{x}{\eta}\right){_2F_1}\left(-j, j+1,2\bigg|\frac{1}{2}\left(1+\frac{x}{\eta}\right)\right)\,,\nonumber\\
\label{Def-p-Q}
&&\mathcal{Q}_j\left(\frac{x}{\eta}\right) = \frac{\sin(\pi j)}{\pi}\,{_2F_1}\left(\frac{j}{2},\frac{j+1}{2}; \frac{3}{2} + j\bigg|\frac{\eta^2}{x^2}\right)\,. 
%\,,\nonumber\\
\eea

The non-singlet isovector quark GPD, denoted as $H^{(-)}_{u-d}(x, \eta, t; \mu)$, represents the difference between the up and down quark distributions,
%. They are represented by
\bea
\label{umdGPDfinal}
&&H^{(-)}_{u-d}(x, \eta, t; \mu) = \nonumber\\
&&\frac{1}{2i}\int_{\mathbb{C}} dj\, \frac{1}{\sin(\pi j)} \,p_j(x,\eta)\,\mathbb{F}_{u-d}^{(-)}(j, \eta, t; \mu)\,.
%,\nonumber\\
\eea
The analytically continued conformal PWs $p_j(x,\eta)$ are given in (\ref{Def-p-allhere}), with support $-\eta\leq x\leq 1$. Similarly, the symmetric gluon GPD reads~\cite{Mueller:2005ed}
\begin{eqnarray}
\label{F(x,eta,t)-MBSymmetricvectorGG}
&&H_g^{(+)}(x, \eta, t; \mu) = \nonumber\\
&&\frac{1}{2i} \int_{\mathbb C} dj\, \frac{(-1)}{\sin(\pi j)} \left( {^g\!p}_j(x, \eta) + {^g\!p}_j(-x, \eta) \right) \mathbb{F}_g^{(+)}(j, \eta, t; \mu),\nonumber\\
\end{eqnarray}
where the gluon PWs are
\bea
\label{Def-p-allGluonhere}
&&{^g\!p}_j(x, \eta) = \nonumber\\
&&\theta(\eta - |x|) \frac{1}{\eta^{j-1}} {^g\!\cal P}_j\left(\frac{x}{\eta}\right) + \theta(x - \eta) \frac{1}{x^{j-1}} {^g\!\cal Q}_j\left(\frac{x}{\eta}\right)\,,\nonumber\\
&&{^g\!\cal P}_j\left(\frac{x}{\eta}\right) =\frac{2^{j-1}\Gamma(3/2+j)}{\Gamma(1/2)\Gamma(j-1)}\nonumber\\ 
&&\times\left(1+\frac{x}{\eta}\right)^2 {_2F_1}\left(-j, j+1, 3 \bigg|\frac{1}{2}\left(1+\frac{x}{\eta}\right)\right)\,,\nonumber\\
&&{^g\!\cal Q}_j\left(\frac{x}{\eta}\right) = -\frac{\sin(\pi j)}{\pi}\,{_2F_1}\left(\frac{j-1}{2}, \frac{j}{2}; \frac{3}{2} + j; \frac{\eta^2}{x^2}\right)\,.
%.\nonumber\\
\eea
Note that the quark and gluon PWs are continuous across $x=\eta$ since $\mathcal{P}_j\left(1\right)=\mathcal{Q}_j\left(1\right)$, and ${^g\!\cal P}_j\left(1\right)={^g\!\cal Q}_j\left(1\right)$.

%%%%%

%%%%%

%\begin{widetext}

%\end{widetext}

%\makeatletter
%\renewcommand{\thesubfigure}{\ifnum\c@subfigure<6 \textcolor{blue}{\textbf{\alph{subfigure}}}\else\ifnum\c@subfigure<12 \textcolor{red}{\textbf{\alph{subfigure}}}\else\textcolor{green}{\textbf{\alph{subfigure}}}\fi\fi}
%\makeatother
\begin{figure}[ht!]
\centering
\subfloat[\label{FUminusDn1}]{%
\includegraphics[height=4cm,width=4cm]{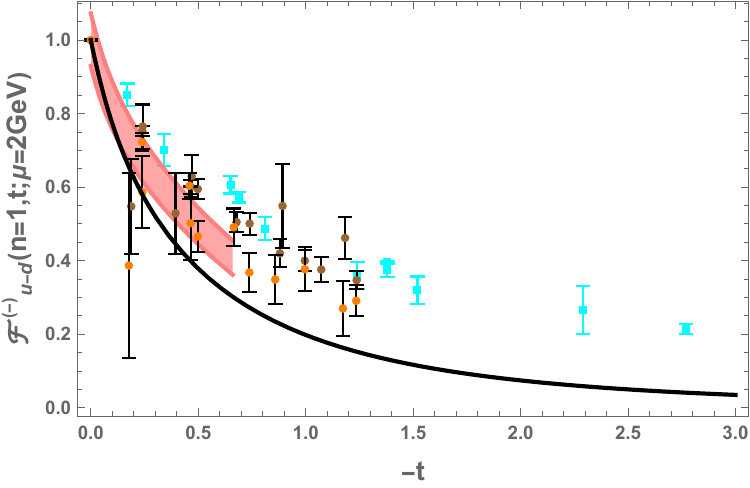}%
}
\subfloat[\label{FUminusDn2}]{%
\includegraphics[height=4cm,width=4cm]{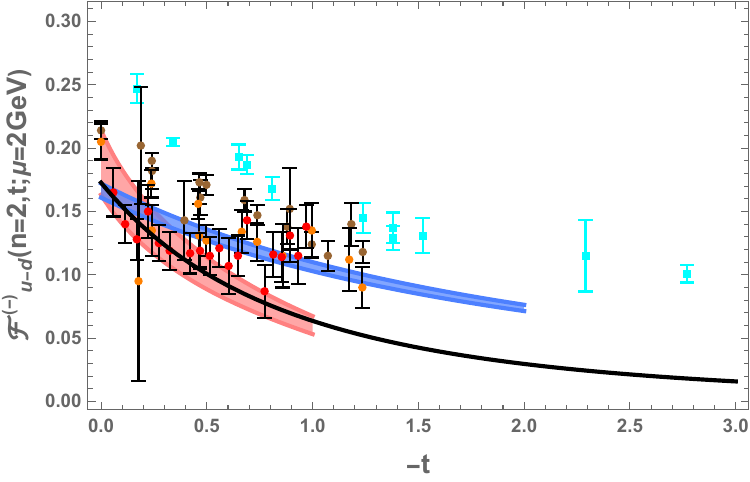}%
}
%\subfloat[\label{FUminusDn3}]{%
%\includegraphics[height=4cm,width=.22\linewidth]{FUminusDn3.pdf}%
%}
%\subfloat[\label{FUminusDn4}]{%
%\includegraphics[height=4cm,width=.22\linewidth]{FUminusDn4.pdf}%
%}
%\subfloat[\label{FUminusDn5}]{%
%\includegraphics[height=4cm,width=.22\linewidth]{FUminusDn5.pdf}%
%}
\\
\subfloat[\label{FUplusDn1}]{%
\includegraphics[height=4cm,width=4cm]{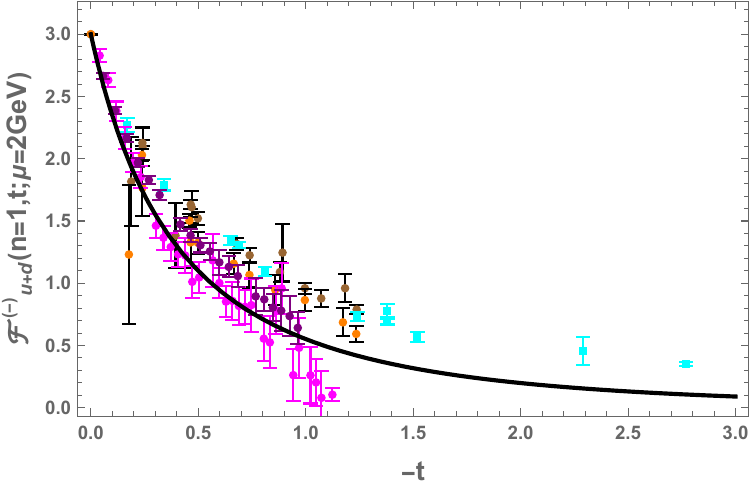}%
}
\subfloat[\label{FUplusDn2}]{%
\includegraphics[height=4cm,width=4cm]{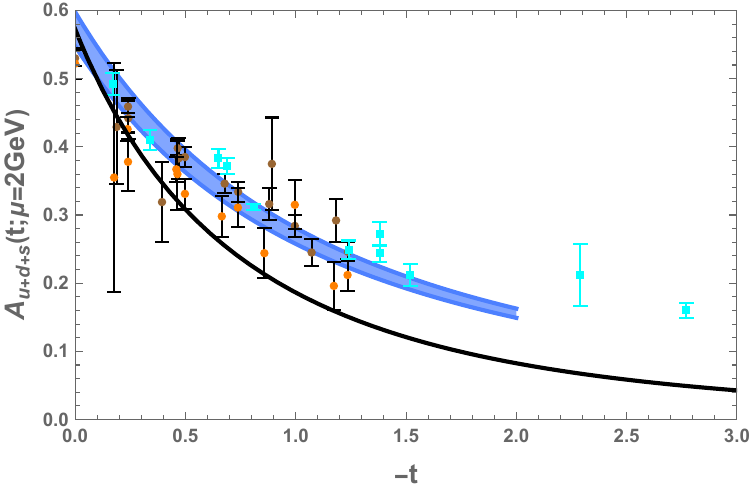}%
}
%\subfloat[\label{DUplusDplusSn2}]{%
%\includegraphics[height=4cm,width=.22\linewidth]{DUplusDplusSn2.pdf}%
%}
%\subfloat[\label{FUplusDn3}]{%
%\includegraphics[height=4cm,width=.22\linewidth]{FUplusDn3.pdf}%
%}
%\subfloat[\label{FUplusDn4}]{%
%\includegraphics[height=4cm,width=.22\linewidth]{FUplusDn4.pdf}%
%}
%\vspace{0.1cm}
\\
%\subfloat[\label{FUplusDn5}]{%
%\includegraphics[height=4cm,width=.22\linewidth]{FUplusDn5.pdf}%
%}
\subfloat[\label{AgluonN2}]{%
\includegraphics[height=4cm,width=4cm]{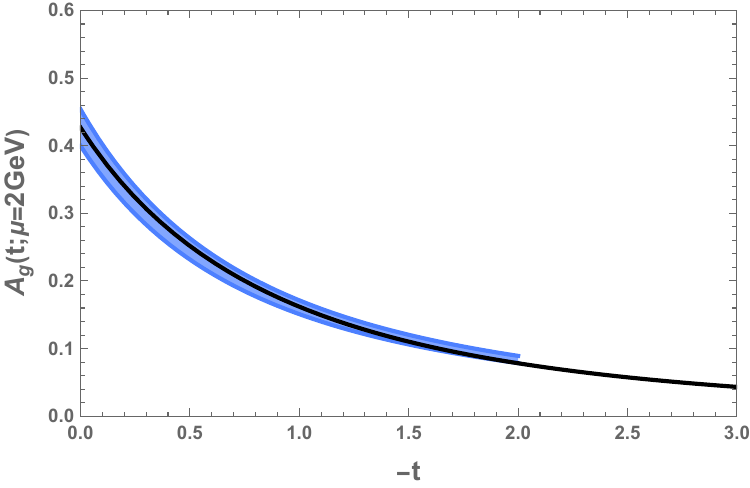}%
}
\subfloat[\label{DgluonN2}]{%
\includegraphics[height=4cm,width=4cm]{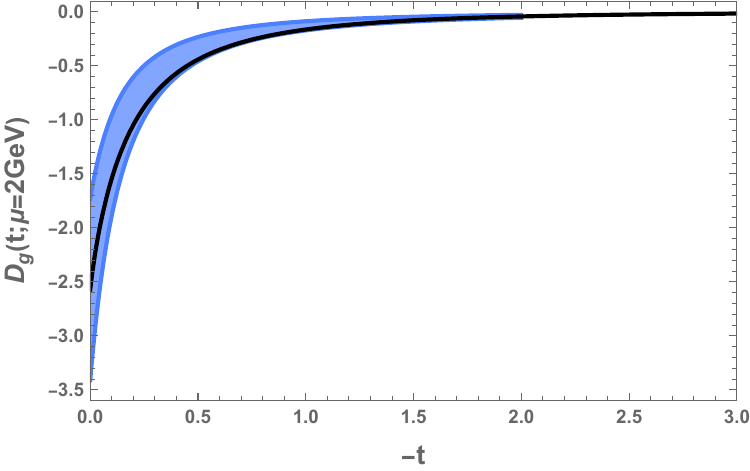}%
}
\caption{Evolved conformal moments $j=n=1,2$ of the quark $u-d$ (a,b), the quark
$u+d$ (c,d), and the gluon for $n=2$ (e,f) GPDs  at the resolution $\mu=2~\text{GeV}$. 
%Moments of $u-d$ quark GPD $H^{u-d}(x,\eta,t;\mu)$ are shown in (\textcolor{blue}{a}-\textcolor{blue}{e}) for $n=1-5$, respectively. Moments of $u+d$ quark GPD $H^{u+d}(x,\eta,t;\mu)$ are depicted in (\textcolor{red}{f}-\textcolor{red}{k}) for $n=1-5$, respectively, where (\usepackage{xcolor}\textcolor{red}{h}) is the quark D-term . Moments of gluon GPD $H_{g}^{(+)}(x,\eta,t;\mu)$ are presented in (\textcolor{green}{l} and \textcolor{green}{m}) for $n=2$, where (\textcolor{green}{m}) is the gluon D-term. 
Our results are illustrated with black curves. The lattice data
are from~\cite{Bhattacharya:2023ays} (Cyan), \cite{LHPC:2007blg} (Orange and Brown), \cite{Lin:2020rxa} (Pink), \cite{Alexandrou:2019ali} (Red), \cite{Alexandrou:2018sjm} (Purple), \cite{Djukanovic:2023beb} (Magenta), and \cite{Hackett:2023rif} (Blue).}
\label{momentsAll}
\end{figure}

%\begin{widetext}
\begin{figure}
%[ht!]
\centering
\subfloat[\label{HUminusDoursAll}]{%
\includegraphics[height=5cm,width=6cm]{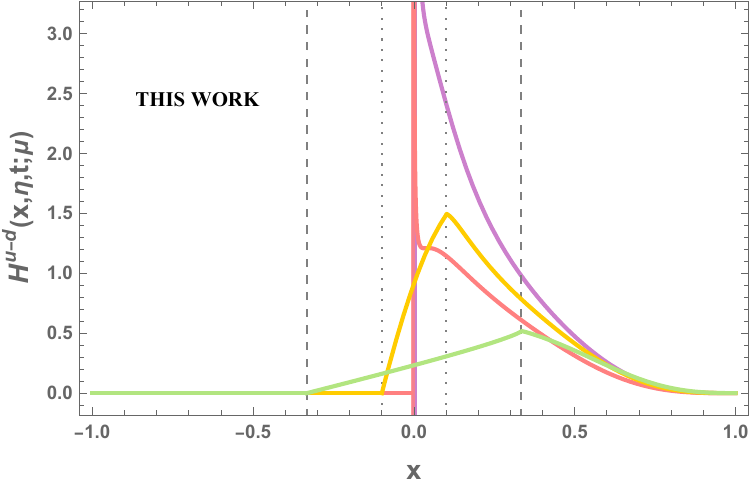}%
}
%\hfill
\\
\subfloat[\label{HUminusDLatticeMarthaLinAll}]{%
\includegraphics[height=5cm,width=6cm]{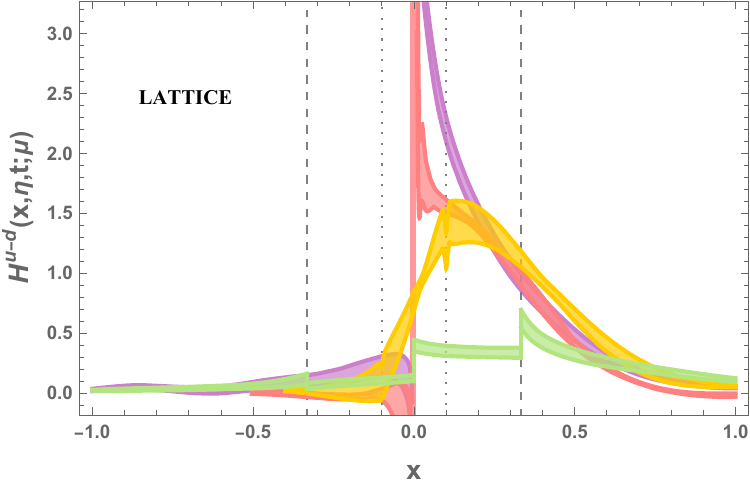}%
}
\caption{(a) Our numerical results (a) for the $u-d$ (isovector) quark GPD in (\ref{umdGPDfinal}) using the conformal moments (\ref{F1jj2}), interpolated from our numerical data all at a resolution $\mu=2\,\rm GeV$: Purple $\eta=0, -t=0$,  Green  $\eta=\frac 13, -t=0.69~\rm GeV^2$, Yellow $\eta=0.1, -t=0.23~\rm GeV^2$,  and Pink for $\eta=0, -t=0.39~\rm GeV^2$. The corresponding lattice results in (b) are from~\cite{Alexandrou:2020zbe} (Purple and Green), from~\cite{Holligan:2023jqh} (Yellow), and from~\cite{Lin:2020rxa} (Pink).}
\label{HUMINUSDGPDs2}
\end{figure}
%\end{widetext}

%\begin{widetext}
\begin{figure}
%[ht!]
\centering
\subfloat[\label{HuAll}]{%
\includegraphics[height=5cm,width=6cm]{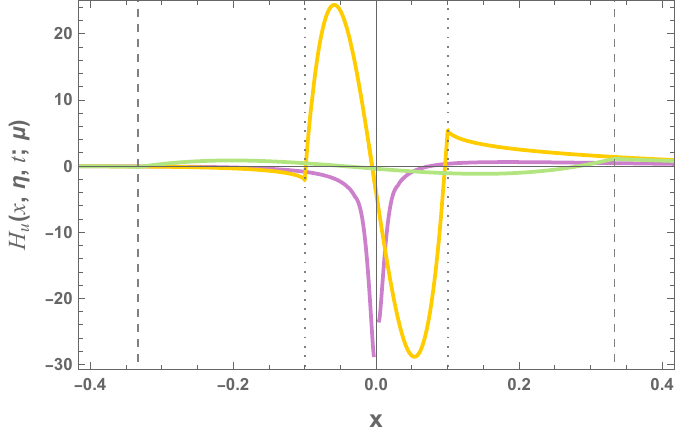}%
}
%\hfill
%\subfloat[\label{HdAll}]{%
%\includegraphics[height=5cm,width=.45\linewidth]{HdAll.pdf}%
%}
\hfill
%\\
\subfloat[\label{gluonHallData}]{%
\includegraphics[height=5cm,width=6cm]{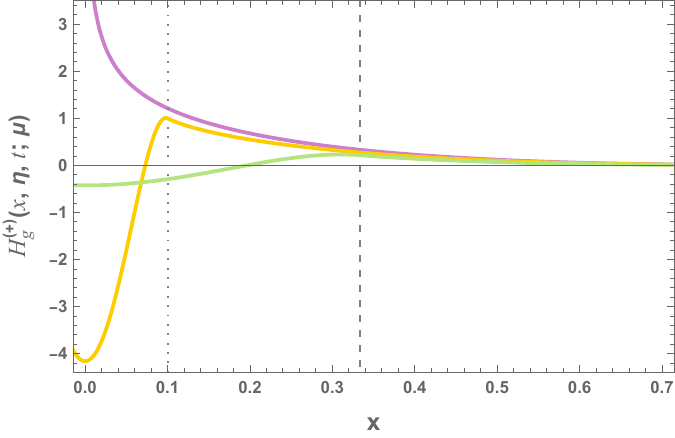}%
}
\caption{Our predictions for the flavor separated $u$-$quark$  and $gluon$ GPDs $H_{u,g}(x,\eta,t;\mu)$ at the resolution $\mu=2\, \rm GeV$:  Purple for $\eta=0,-t=0$,  Green for $\eta=\frac 13, -t=0.69~\rm GeV^2$, and Yellow for $\eta=0.1, -t=0.23~\rm GeV^2$.}
\label{HUD}
\end{figure}

\textit{\bf String-based moment parametrization}.---The holographic string approach offers a parametrization for the conformal moments of both non-singlet $(-)$ and singlet $(+)$ quark GPD combinations at the resolution  $\mu = \mu_0$. 

The holographic parametrization of the conformal moments of the singlet (sea) quark GPDs at $\mu=\mu_0$ is represented by the sum of the sea quark spin-j A-form factor and skewness-dependent spin-j D-form factor as  
\bea\label{seaMomentFinal}
&&\sum_{q=1}^{N_f}\,\mathbb{F}_{q}^{(+)} (j,\eta, t ; \mu_0) =\nonumber\\
&&\sum_{q=1}^{N_f}\,\mathcal{F}_{q}^{(+)}(j,t;\mu_0)+\mathcal{F}_{q\eta}^{(+)}(j,\eta,t;\mu_0)\,,
\eea
for even $j=2,4,\cdots$, where we have defined
%\begin{widetext}
\bea\label{singletF1jj}
&&\sum_{q=1}^{N_f}\,\mathcal{F}_{q}^{(+)}(j,t;\mu_0) = \int_{0}^{1}\,dx\,\sum_{q=1}^{N_f}
\frac{q^{(+)}(x;\mu_0)}{x^{1-j+\alpha^{\prime}_{(+)}t}}\,,\nonumber\\\eea
with the input singlet quark PDF 
\bea
&&\sum_{q=1}^{N_f}\,q^{(+)}(x;\mu_0)=\sum_{q=1}^{N_f}\,q(x;\mu_0)+\bar q(x;\mu_0)\,.
\eea
%at $\mu=\mu_0$.
%and $a_q\equiv -\alpha^{\prime}_q t$. 
The skewness or
$\eta$-dependent 
spin-j D-form factor, satisfying the polynomiality condition at any skewness, is given by
\bea\label{singletF1etaKj22}
&&\sum_{q=1}^{N_f}\,\mathcal{F}_{q\eta}^{(+)}(j,\eta,t;\mu_0)=\left(\hat{d}_{j}(\eta,t)-1\right)\nonumber\\
&&\times \left[\sum_{q=1}^{N_f}\,\mathcal{F}_{q}^{(+)}(j,t;\mu_0)-\mathcal{F}_{q,s}^{(+)}(j,t;\mu_0)\right]
%\,,\nonumber\\
\eea
where
\bea
\label{singletF1etaSj22}
\sum_{q=1}^{N_f}\,\mathcal{F}_{q,s}^{(+)}(j,t;\mu_0)&\equiv&\sum_{q=1}^{N_f}\,\mathcal{F}_{q}^{(+)}(j,t;\mu_0,\alpha^{\prime}_{(+)}\rightarrow \alpha^{\prime}_{(+),s})\,,\nonumber\\
\eea
and the model-independent (universal), $\eta$-dependent coefficient function is given in terms of hypergeometric function~\cite{Nishio:2014eua,Mamo:2022jhp} 
\bea\label{singletetapoly232}
&&\hat{d}_{j}(\eta,t)-1=\nonumber\\
&&\, _2F_1\left(-\frac{j}{2},-\frac{j-1}{2};\frac{1}{2}-j;\frac{4 m_N^2}{-t}\times\eta ^2\right)-1\,,
%\,.\nonumber\\
\eea
which is a finite polynomial in $\eta$ of degree $j=2m$ for $m\geq 1$, is due to the Lorentz invariance of the t-channel spin-j resonance (string) exchanges in AdS.
Note that the spin-2 A-form factor of the quark gravitational form factor of the proton is 
$
\sum_{q=1}^{N_f}\,A_q(t;\mu_0)=\sum_{q=1}^{N_f}\,\mathcal{F}_{q}^{(+)}(2,t;\mu_0)\,,
$
and the D-form factor (or the D-term) of the quark gravitational form factor of the proton is 
$
\eta^2 \sum_{q=1}^{N_f}\,D_q(t;\mu_0)=\sum_{q=1}^{N_f}\,\mathcal{F}_{q\eta}^{(+)}(2,\eta,t;\mu_0)\,.
$

The non-singlet quark conformal moment is  represented by the valence quark spin-j A-form factor~\cite{Mamo:2022jhp} 
\bea\label{valenceMomentFinal}
&&\mathbb{F}_{q}^{(-)} (j,\eta, t ; \mu_0) =\mathcal{F}_{q}^{(-)}(j,t;\mu_0)
\,,
\eea
for odd $j=1,3,\cdots$, which is tied to the valence (non-singlet) quark PDF through
%\begin{widetext}
\bea\label{F1jj}
\mathcal{F}_{q}^{(-)}(j,t;\mu_0) = \int_{0}^{1}\,dx\,
\frac{q_v(x;\mu_0)}{x^{1-j+\alpha^{\prime}_{(-)}t}}\,,
\eea
with $q_v(x;\mu_0)=q(x;\mu_0)-\bar q(x;\mu_0)=H^{(-)}_q (x, \eta=0, t=0 ; \mu_0)$ at $\mu=\mu_0$. We can also write
\bea\label{F1jj2}
\mathbb{F}_{u\pm d}^{(-)} (j,\eta, t ; \mu_0) &=&\mathcal{F}_{u\pm d}^{(-)}(j,t;\mu_0)\nonumber\\
&=&\int_{0}^{1}\,dx\,
\frac{u_v(x;\mu_0)\pm d_v(x;\mu_0)}{x^{1-j+\alpha^{\prime}_{u\pm d}t}}\,,\nonumber\\
\eea
assuming $\bar{u}(x,\mu_0)=\bar{d}(x,\mu_0)$. We can fix $\alpha_{u\pm d}'$, using the experimental Dirac electromagnetic form factor of the proton ($F_{1p}(t)$) and neutron ($F_{1n}(t)$) from their valence combination
$
\mathcal{F}_{u+d}^{(-)}(1,t;\mu_0)=3\left(F_{1p}(t)+F_{1n}(t)\right)\,,
$
and the isovector combination
$
\mathcal{F}_{u-d}^{(-)}(1,t;\mu_0)=F_{1p}(t)-F_{1n}(t)\,.
$
We specifically use the experimental Dirac electromagnetic form factor of the nucleon from~\cite{Mamo:2021jhj}. In (\ref{valenceMomentFinal}) and (\ref{F1jj2}), we have ignored the valence (non-singlet) quark skewness-dependent spin-j D-form factor which is vanishing (since the pressure distribution inside the nucleon, as determined by the D-form factor, is flavor or electric charge independent).

Note that for $j=1$ and low-x, the integrands asymptote 
\bea
\frac{q_v(x;\mu_0)}{x^{\alpha_{(-)}^{\prime} t}}\rightarrow 
\bigg(\frac 1x\bigg)^{\alpha_{(-)}(0)+\alpha_{(-)}^{\prime} t}\,,
\eea
which is the expected open-string Regge limit, with
intercept $\alpha_{(-)}(0)$ and slope $\alpha_{(-)}^{\prime}$. Also note that (\ref{singletF1jj}) and (\ref{F1jj})
can be derived holographically using t-channel string exchange in deformed AdS~\cite{Mamo:2022jhp}, however the result is dependent on the choice of the particular deformation of the AdS background metric such as hard-wall, and soft-wall AdS spaces. Here we will fix them
using the empirical PDFs.

The holographic parametrization of the conformal moments of the gluon GPD at finite skewness and resolution  $\mu=\mu_0$, is represented by the sum of the gluon spin-j A-form factor and skewness-dependent spin-j D-form factor as~\cite{Mamo:2022jhp}  
\bea\label{gluonMomnetFinal}
\mathbb{F}_g^{(+)} (j,\eta, t ; \mu_0)=\mathcal{A}_g(j, t;\mu_0)+\mathcal{D}_{g\eta}(j,\eta,t;\mu_0)
%\,,\nonumber\\
\eea 
for even $j=2,4,...$, where we have defined
%\begin{widetext}
\bea\label{Ajj}
&&\mathcal{A}_{g}(j,t;\mu_0) = \int_{0}^{1}\,dx\,
\frac{xg(x;\mu_0)}{x^{2-j+\alpha_T^{\prime}t}}\,.
\eea
Again we note that for $j=2$ and low-x, the integrand asymptotes 
\bea
\frac{xg(x;\mu_0)}{x^{\alpha_T^{\prime} t}}\rightarrow 
\bigg(\frac 1x\bigg)^{\alpha_T(0)+\alpha_T^{\prime} t}\,,
\eea
which is the expected closed-string Regge limit, with
intercept $\alpha_T(0)$ and slope $\alpha^{\prime}_{T}(0)$. Although (\ref{Ajj})
can be derived holographically~\cite{Mamo:2022jhp}, here we choose to fix it 
again using the empirical PDFs. 

The skewness or
$\eta$-dependent spin-j D-form factors, satisfying the polynomiality condition at any skewness, are given by 
\bea\label{DKj22}
&&\mathcal{D}_{g\eta}(j,\eta,t;\mu_0)=\nonumber\\
&&\left(\hat{d}_{j}(\eta,t)-1\right)\times \left[\mathcal{A}_{g}(j,t;\mu_0)-\mathcal{A}_{gS}(j,t;\mu_0)\right]\,,\nonumber\\
\eea
where
\bea
\label{ASj22}
\mathcal{A}_{gS}(j,t;\mu_0)&\equiv&\mathcal{A}_{g}(j,t;\mu_0,\alpha^{\prime}_T\rightarrow \alpha^{\prime}_S)\,,
\eea
with the Regge slope parameter $\alpha^{\prime}_S$ to be fixed by the gluon gravitational form factor of the proton, and $\hat d_j(\eta, t)-$ given by (\ref{singletetapoly232}). 
Note that the spin-2 A-form factor of the gluonic gravitational form factor of the proton is given by
$
A_g(t;\mu_0)=\mathcal{A}_{g}(2,t;\mu_0)\,,
$
and the D-form factor (or the D-term) of the gluonic gravitational form factor of the proton is given by 
$
\eta^2 D_g(t;\mu_0)=\mathcal{D}_{g\eta}(2,\eta,t;\mu_0)\,.
$

Recently, the $J/\Psi-007$ collaboration at JLab~\cite{Duran:2022xag}, using our holographic prediction for near-threshold $J/\Psi$ expressed in terms of $\mathbb{F}_g^{(+)} (2,\eta, t ; \mu_0)$ as defined in (\ref{gluonMomnetFinal}) and detailed in~\cite{Mamo:2019mka,Mamo:2022eui}, has successfully extracted the gluonic spin-2 moment. The empirical results align with both holography and lattice QCD~\cite{Hackett:2023rif}, providing strong support for the current approach.

%%%%%%%%%%%%%%%%%555
\textit{\bf Results}.---We initialized the conformal moments of quark and gluon GPDs using  the empirical quark and gluon PDFs at a low resolution point, $\mu=\mu_0$. Employing the holographic parametrization and empirical Regge slope parameters from the electromagnetic and gravitational form factors of the proton, as specified in Table~\ref{table1}, we adopted the leading-order Martin-Stirling-Thorne-Watt (MSTW) 2008 PDF sets at $\mu_0=1~\text{GeV}$, assuming $\Delta(x;\mu_0)=\bar{d}(x;\mu_0)-\bar{u}(x;\mu_0)\approx 0$. For more details on the numerical results see the expanded version~\cite{Mamo:2024vjh}.

Our findings, shown as black-solid lines in Fig.\ref{momentsAll}, compare the evolved lowest moments ($n=1,2,...,5$) for nucleon $u\pm d$ quark and gluon GPDs at $\mu=2~\text{GeV}$ against recent lattice data, shown as data points or colored spreads. Additionally, in Fig.~\ref{HUMINUSDGPDs2}a, we present the full GPDs for the nucleon non-singlet $u-d$ quark across various skewness $\eta$ and Mandelstam $t$ values, to be compared to the lattice QCD result in Fig.~\ref{HUMINUSDGPDs2}b.

In Fig.~\ref{HUD}, we  show the flavor-separated GPDs for the nucleon $u, d, g$ quarks (anti-quarks) and gluon, excluding the strange quark contribution and approximating $H_{u+d+s}^{(+)}(x)\approx H_{u+d}^{(+)}(x)$, where we have  suppressed the arguments $\eta$, $t$, and $\mu$ for clarity. In addition, the quark GPDs (in the positive-x (+x) region) and anti-quark GPDs (in the negative-x (-x) region) are defined by $H_q(\pm x)=\frac{1}{2}\left(H_q^{(-)}(x)\pm H_q^{(+)}(x)\right)$, respectively. We have also defined $H_{u+d}(\pm x)\equiv\frac{1}{2}\left(H_{u+d}^{(-)}(x)\pm H_{u+d}^{(+)}(x)\right)$, and $H_{u-d}(\pm x)\equiv H_{u-d}^{(-)}(\pm x)$ which can be combined for the flavor separation. Note that these GPDs converge to their respective PDFs at $\eta=t=0$ with $q(x)=H_q(+x)$, and $\bar q(x)=-H_q(-x)$. %We have also defined $H_{u+d}(\pm x,\eta,t)\equiv\frac{1}{2}\left(H_{u+d}^{(-)}\pm H_{u+d}^{(+)}\right)$ and $H_{u-d}(\pm x,\eta,t)\equiv H_{u-d}^{(-)}(\pm x,\eta,t)$ so that $H_{u}(\pm x,\eta,t)=\frac{1}{2}\left(H_{u+d}(\pm x,\eta,t) + H_{u-d}(\pm x,\eta,t)\right)$ and $H_{d}(\pm x,\eta,t)=\frac{1}{2}\left(H_{u+d}(\pm x,\eta,t) - H_{u-d}(\pm x,\eta,t)\right)$. 

%%%%%%%%%%%%%%%%%%%555
\begin{table}
%[ht!]
\caption{Regge slope parameters for the string-based parametrization of conformal moments.}
\label{table1}
\centering
\begin{tabular}{cc}
\hline \hline
Parameter & Value ($\rm GeV^{-2}$) \\
\hline
$\alpha^{\prime}_T$ & 0.627 \\
$\alpha^{\prime}_S$ & 4.277 \\
$\alpha^{\prime}_{(+)}=\alpha^{\prime}_{(-)}=\alpha^{\prime}_{u+d}$ & 0.891 \\
$\alpha^{\prime}_{u-d}$ & 1.069 \\
$\alpha^{\prime}_{(+),s}$ & 1.828 \\
\hline \hline
\end{tabular}
\end{table}

\textit{\bf Summary and Outlook}.---Our string-based parametrization extends our recent work on nucleon GPDs from the ERBL regime at low-x to the DGLAP regime at high-x~\cite{Mamo:2022jhp}.
The spin-j conformal moments encode the GPDs hadronic content, described in the holographic  framework as Reggeized exchanges with few parameters, i.e. the Regge intercepts and slopes.  Our isovector quark GPDs are in overall agreement with the available lattice data. Our new results for the gluon, valence and sea GPDs are yet to be tested both on the lattice and to future measurements. We plan to extend this analysis to polarized and helicity flip GPDs. On a broader note, our string based parametrization, which enforces the requirements of polynomiality, symmetry and analyticity, addresses the deconvolution challenge, and promises to help significantly the global data analysis of GPDs at JLab and the upcoming EIC.

\textit{\bf  Acknowledgments}.---
K.M. is supported by U.S. DOE Grant No. DE-FG02-04ER41302, and thanks Konstantinos Orginos and Christian Weiss for discussions and hospitality at JLab where part of this work was carried out. I.Z. is supported by the  U.S. DOE Grant  No. DE-FG-88ER40388.
This research is also supported in part within the framework of the Quark-Gluon Tomography (QGT) Topical Collaboration, under contract no. DE-SC0023646.

\bibliography{GPDs_DVCS}

\end{document}